\documentclass[runningheads]{llncs}
\usepackage[T1]{fontenc}
\usepackage{amsfonts}
\usepackage{multirow}
\usepackage{booktabs}
\usepackage{makecell}
\usepackage{graphicx}
\usepackage{algorithm}
\usepackage{amsmath}
\usepackage{setspace}
\usepackage{algpseudocode}
\begin{document}
\title{MedGen3D: A Deep Generative Framework for Paired 3D Image and Mask Generation}
\titlerunning{MedGen3D: Paired 3D Image and Mask Generation}
%
\author{Kun Han\inst{1} \and
Yifeng Xiong\inst{1} \and
Chenyu You\inst{2} \and
Pooya Khosravi\inst{1} \and
Shanlin sun\inst{1} \and
Xiangyi Yan\inst{1} \and
James Duncan\inst{2} \and
Xiaohui Xie\inst{1}}
\authorrunning{K. Han et al.}

\institute{University of California, Irvine \and Yale University}
\maketitle              

\vspace{-2em}
\begin{abstract}
Acquiring and annotating sufficient labeled data is crucial in developing accurate and robust learning-based models, but obtaining such data can be challenging in many medical image segmentation tasks. One promising solution is to synthesize realistic data with ground-truth mask annotations. However, no prior studies have explored generating complete 3D volumetric images with masks.
In this paper, we present MedGen3D, a deep generative framework that can generate paired 3D medical images and masks. First, we represent the 3D medical data as 2D sequences and propose the Multi-Condition Diffusion Probabilistic Model (MC-DPM) to generate multi-label mask sequences adhering to anatomical geometry. Then, we use an image sequence generator and semantic diffusion refiner conditioned on the generated mask sequences to produce realistic 3D medical images that align with the generated masks.
Our proposed framework guarantees accurate alignment between synthetic images and segmentation maps. Experiments on 3D thoracic CT and brain MRI datasets show that our synthetic data is both diverse and faithful to the original data, and demonstrate the benefits for downstream segmentation tasks. 
We anticipate that MedGen3D's ability to synthesize paired 3D medical images and masks will prove valuable in training deep learning models for medical imaging tasks.


\end{abstract}
\section{Introduction}
\label{intro}

In medical image analysis, the availability of a substantial quantity of accurately annotated 3D data is a prerequisite for achieving high performance in tasks like segmentation and detection \cite{ronneberger2015u,hatamizadeh2022unetr,tang2022self,chen2021deep,yan2022after,you2022bootstrapping,you2022class,you2022mine,you2023rethinking}. This, in turn, leads to more precise diagnoses and treatment plans. However, obtaining and annotating such data presents many challenges, including the complexity of medical images, the requirement for specialized expertise, and privacy concerns.

Generating realistic synthetic data presents a promising solution to the above challenges as it eliminates the need for manual annotation and alleviates privacy risks. However, most prior studies \cite{han2018gan,baur2018melanogans,bermudez2018learning,abhishek2019mask2lesion,abbasi20204d,kim2021synthesis,sun2022hierarchical,you2022bootstrapping,you2022mine,you2023rethinking} have focused on 2D image synthesis, with only a few generating corresponding segmentation masks.
For instance, ~\cite{guibas2017synthetic} uses dual generative adversarial networks (GAN) \cite{goodfellow2020generative,you2022class} to synthesize 2D labeled retina fundus images, while \cite{fernandez2022can} combines a label generator \cite{rombach2022high} with an image generator \cite{park2019semantic} to generate 2D brain MRI data. More recently, \cite{subramaniam2022generating} uses WGAN \cite{arjovsky2017wasserstein} to generate small 3D patches and corresponding vessel segmentations. 

However, there has been no prior research on generating whole 3D volumetric images with the corresponding segmentation masks.
Generating 3D volumetric images with corresponding segmentation masks faces two major obstacles. First, directly feeding entire 3D volumes to neural networks is impractical due to GPU memory constraints, and downsizing the resolution may compromise the quality of the synthetic data. Second, treating the entire 3D volume as a single data point during training is suboptimal because of the limited availability of annotated 3D data. Thus, innovative methods are required to overcome these challenges and generate high-quality synthetic 3D volumetric data with corresponding segmentation masks.



We propose MedGen3D, a novel diffusion-based deep generative framework that generates paired 3D volumetric medical images and multi-label masks. Our approach treats 3D medical data as sequences of slices and employs an autoregressive process to sequentially generate 3D masks and images. In the first stage, a Multi-Condition Diffusion Probabilistic Model (MC-DPM) generates mask sequences by combining conditional and unconditional generation processes. Specifically, the MC-DPM generates mask subsequences (i.e., several consecutive slices) at any position directly from random noise or by conditioning on existing slices to generate subsequences forward or backward. Given that medical images have similar anatomical structures, slice indices serve as additional conditions to aid the mask subsequence generation. In the second stage, we introduce a conditional image generator with a seq-to-seq model from \cite{wang2018video} and a semantic diffusion refiner. By conditioning on the mask sequences generated in the first stage, our image generator synthesizes realistic medical images aligned with masks while preserving spatial consistency across adjacent slices.

The main contributions of our work are as follows:
1) Our proposed framework is the \emph{first} to address the challenge of synthesizing complete 3D volumetric medical images with their corresponding masks;
2) we introduce a multi-condition diffusion probabilistic model for generating 3D anatomical masks with high fidelity and diversity;
3) we leverage the generated masks to condition an image sequence generator and a semantic diffusion refiner, which produces realistic medical images that align accurately with the generated masks; and
4)  we present experimental results that demonstrate the fidelity and diversity of the generated 3D multi-label medical images, highlighting their potential benefits for downstream segmentation tasks.


\vspace{-0.5em}
\section{Preliminary}
\vspace{-0.5em}
\subsection{Diffusion Probabilistic Model}
A diffusion probabilistic model (DPM) \cite{ho2020denoising} is a parameterized Markov chain of length T, which is designed to learn the data distribution $p(X)$. DPM builds the Forward Diffusion Process (FDP) to get the diffused data point $X_t$ at any time step $t$ by $q\left(X_t \mid X_{t-1}\right)=\mathcal{N}\left(X_t ; \sqrt{1-\beta_t} X_{t-1}, \beta_t I\right)$, with $X_0 \sim q(X_0)$ and $p(X_T) = \mathcal{N}\left(X_T ; 0, I\right)$. Let $\alpha_t=1-\beta_t $ and $ \bar{\alpha}_t=\prod_{s=1}^t\left(1-\beta_s\right)$, Reverse Diffusion Process (RDP) is trained to predict the noise added in the FDP by minimizing:
\vspace{-0.8em}
\begin{equation}
    Loss(\theta)=\mathbb{E}_{X_0 \sim q\left(X_0\right), \epsilon \sim \mathcal{N}(0, I), t}\left[\left\|\epsilon-\epsilon_\theta\left(\sqrt{\bar{\alpha}_t} X_0+\sqrt{1-\bar{\alpha}_t} \epsilon, t\right)\right\|^2\right],
\vspace{-0.5em}
\end{equation}
where $\epsilon_\theta$ is predicted noise and $\theta$ is the model parameters.
\vspace{-1em}

\subsection{Classifier-free Guidance}
\label{cf_guidance}
Samples from conditional diffusion
models can be improved with classifier-free guidance \cite{ho2022classifierfree} by setting the condition $c$ as $\emptyset$ with probability $p$. During sampling, the output of the model is extrapolated further in the direction of $\epsilon_\theta\left(X_t \mid c\right)$ and away from $\epsilon_\theta\left(X_t \mid \emptyset\right)$ as follows:
\vspace{-0.8em}
\begin{equation}
    \hat{\epsilon}_\theta\left(X_t \mid c\right)=\epsilon_\theta\left(X_t \mid \emptyset\right)+s \cdot\left(\epsilon_\theta\left(X_t \mid c\right)-\epsilon_\theta\left(X_t \mid \emptyset\right)\right),
\vspace{-0.8em}
\end{equation}
where $\emptyset$ represents a null condition and $s \geq 1$ is the guidance scale. 
\vspace{-0.5em}

\section{Methodology}
\vspace{-0.5em}
We propose a sequential process to generate complex 3D volumetric images with masks, as illustrated in Figure 1. 
The first stage generates multi-label segmentation, and the second stage performs conditional medical image generation. 
The details will be presented in the following sections.
\vspace{-1.5em}
\begin{figure}
\includegraphics[width=\textwidth]{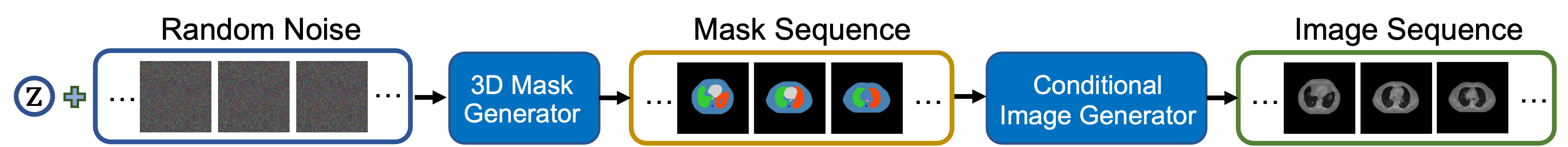}
\caption{Overview of the proposed \textbf{MedGen3D}, including a 3D mask generator to autoregressively generate the mask sequences starting from a random position $z$, and a conditional image generator to generate 3D images conditioned on generated masks.} \label{fig1}
\end{figure}
\vspace{-3em}
\subsection{3D Mask Generator}

Due to the limited annotated real data and GPU memory constraints, directly feeding the entire 3D volume to the network is impractical. Instead, we treat 3D medical data as a series of subsequences. To generate an entire mask sequence, an initial subsequence of $m$ consecutive slices is \textbf{unconditionally} generated from random noise. Then the subsequence is expanded \textbf{forward} and \textbf{backward} in an autoregressive manner, conditioned on existing slices.

\begin{figure}[ht]
\includegraphics[width=\textwidth]{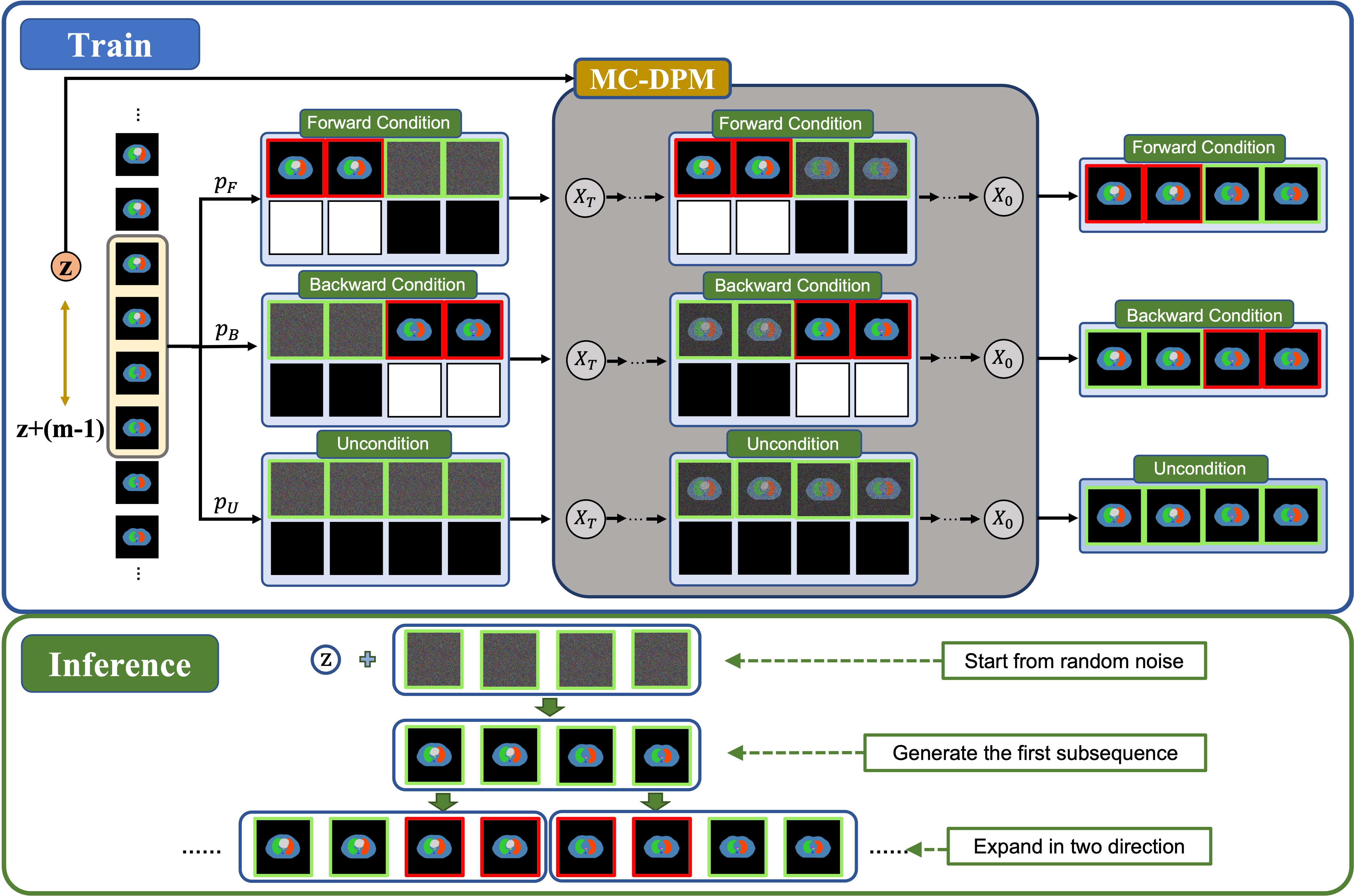}
\vspace{-2em}
\caption{Proposed 3D mask generator. Given target position $z$, MC-DPM is designed to generate mask subsequences (length of $m$) for specific region,  unconditionally or conditioning on first or last $n$ slices, according to the pre-defined probability $p^C \in \{{p_F,p_B,p_U}\}$. Binary indicators are assigned to slices to signify the conditional slices. We ignore the binary indicators in the inference process for clear visualization with red outline denoting the conditional slices and green outline denoting the generated slices.}  \label{fig:mcdpm}
\vspace{-1.5em}
\end{figure}

Inspired by classifier-free guidance in Section \ref{cf_guidance}, we propose a general Multi-Condition Diffusion Probabilistic Model (MC-DPM) to unify all three conditional generations (unconditional, forward, and backward). As shown in Fig. \ref{fig:mcdpm}, MC-DPM is able to generate mask sequences directly from random noise or conditioning on existing slices.


Furthermore, as 3D medical data typically have similar anatomical structures, slices with the same relative position roughly correspond to the same anatomical regions. Therefore, we can utilize the relative position of slices as conditions to guide the MC-DPM in generating subsequences of the target region and control the length of generated sequences. 

\noindent{\bfseries Train:} For a given 3D multi-label mask $M \in \mathbb{R}^{D \times H \times W}$, subsequneces of $m$ consecutive slices are selected as $\{M_z, M_{z+1}, \ldots, M_{z+(m-1)}\}$, with $z$ as the randomly selected starting indices. For each subsequence, we determine the conditional slices $X^C \in \{\mathbb{R}^{n \times H \times W}, \emptyset\}$ by selecting either the first or the last $n$ slices, or no slice, based on a probability $p^C \in \{{p_{Forward},p_{Backward},p_{Uncondition}}\}$. The objective of the MC-DPM is to generate the remaining slices, denoted as $X^P \in \mathbb{R}^{(m-\text{len}(X^C)) \times H \times W}$. 

To incorporate the position condition, we utilize the relative position of the subsequence $\tilde{z}=z/D$, where $z$ is the index of the subsequence's starting slice. Then we embed the position condition and concatenate it with the time embedding to aid the generation process. We also utilize a binary indicator for each slice in the subsequence to signify the existence of conditional slices.

The joint distribution of reverse diffusion process (RDP) with the conditional slices $X^C$ can be written as: 
\vspace{-1em}
\begin{equation}
    p_\theta(X_{0: T}^P | X^C,\tilde{z})=p(X_T^P) \prod_{t=1}^T p_\theta(X_{t-1}^P \mid X_t^P, X^C, \tilde{z}).
\vspace{-1em}
\end{equation}
where $p(X_T^P) = \mathcal{N}\left(X_T^P ; 0, I\right)$, $\tilde{z}=z / D$ and $p_\theta$ is the distribution parameterized by the model. 

Overall, the model will be trained by minimizing the following loss function, with $X_t^P = \sqrt{\bar{\alpha}_t} X_0^P+\sqrt{1-\bar{\alpha}_t} \epsilon$:
\vspace{-0.5em}
\begin{equation}
    \operatorname{Loss}(\theta)=\mathbb{E}_{X_0 \sim q(X_0), \epsilon \sim \mathcal{N}\left(0, I\right), p^C, z, t}\left[\left\|\epsilon-\epsilon_\theta\left(X_t^P, X^{C}, z, t\right)\right\|^2\right].
\vspace{-0.5em}
\end{equation}

\noindent{\bfseries Inference:} During inference, MC-DPM first generates a subsequence of $m$ slices from random noise given a random location $z$. The entire mask sequence can then be generated autoregressively by expanding in both directions, conditioned on the existing slices, as shown in Figure \ref{fig:mcdpm}. Please refer to the \textbf{Supplementary} for a detailed generation process and network structure. 

\vspace{-1em}

\subsection{Conditional Image Generator}
\begin{figure}
\vspace{-2em}
\includegraphics[width=\textwidth]{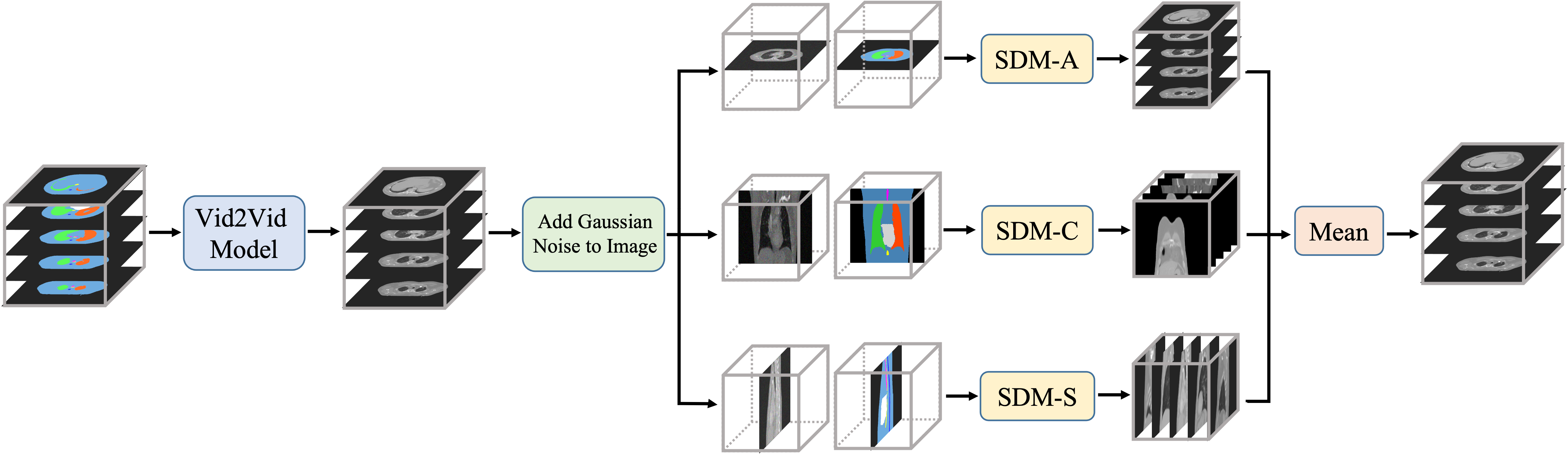}
\vspace{-2.5em}
\caption{Image Sequence Generator. Given the generated 3D mask, the initial image is generated by Vid2Vid model sequentially. To utilize the semantic diffusion model (SDM) to refine the initial result, we first apply small steps (10 steps) noise, and then use three SDMs to refine. The final result is the mean 3D images from 3 different views (Axial, Coronal, and Sagittal), yielding significant improvements over the initially generated image.}  \label{fig:imggen}
\vspace{-1.5em}
\end{figure}

In the second step, we employ a sequence-to-sequence method to generate medical images conditioned on masks, as shown in Figure \ref{fig:imggen}. 

\noindent{\bfseries Image Sequence Generator:} In the sequence-to-sequence generation task, new slice is the combination of the warped previous slice and newly generated texture,  weighted by a continuous mask \cite{wang2018video}. We utilize Vid2Vid \cite{wang2018video} as our image sequence generator. We train Vid2Vid with its original loss, which includes GAN loss on multi-scale images and video discriminators, flow estimation loss, and feature matching loss.

\noindent{\bfseries Semantic Diffusion Refiner:} Despite the high cross-slice consistency and spatial continuity achieved by vid2vid, issues such as blocking, blurriness and suboptimal texture generation persist. Given that diffusion models have been shown to generate superior images \cite{dhariwal2021diffusion}, we propose a semantic diffusion refiner utilizing a diffusion probabilistic model to refine the previously generated images. 

For each of the 3 different views, we train a semantic diffusion model (SDM), which takes 2D masks and noisy images as inputs to generate images aligned with input masks. During inference, we only apply small noising steps (10 steps) to the generated images so that the overall anatomical structure and spatial continuity are preserved. After that, we refine the images using the pre-trained semantic diffusion model. The final refined 3D images are the mean results from 3 views. Experimental results show an evident improvement in the quality of generated images with the help of semantic diffusion refiner.

\section{Experiments and Results}
\subsection{Datasets and Setups}
\noindent{\bfseries Datasets:} We conducted experiments on the thoracic site using three thoracic CT datasets and the brain site with two brain MRI datasets. For both generative models and downstream segmentation tasks, we utilized the following datasets:
\vspace{-0.7em}
\begin{itemize}
    \item SegTHOR \cite{lambert2020segthor}: 3D thorax CT scans (25 training, 5 validation, 10 testing);
    \item OASIS \cite{marcus2007open}: 3D brain MRI T1 scans (40 training, 10 validation, 10 testing);
\end{itemize}
\vspace{-0.7em}
For the downstream segmentation task only and the transfer learning, we utilized 10 fine-tuning, 5 validation, and 10 testing scans from each of the 3D thorax CT datasets of StructSeg-Thorax \cite{structseg} and Public-Thor \cite{chen2021deep}, as well as the 3D brain MRI T1 dataset from ADNI \cite{ADNI}.

%

\noindent{\bfseries Implementation:} For thoracic datasets, we crop and pad CT scans to ($96 \times 320 \times 320$). The annotations of six organs (left lung, right lung, spinal cord, esophagus, heart, and trachea) are examined by an experienced radiation oncologist. We also include a body mask to aid in the image generation of body regions. For brain MRI datasets, we use Freesurfer \cite{fischl2012freesurfer} to get segmentations of four regions (cortex, subcortical gray matter, white matter, and CSF), and then crop the volume to ($192 \times 160 \times 160$). We assign discrete values to masks of different regions or organs for both thoracic and brain datasets and then combine them into one 3D volume. When synthesizing mask sequences, we resize the width and height of the masks to $128 \times 128$ and set the length of the subsequence $m$ to 6. We use official segmentation models provided by MONAI\cite{cardoso2022monai} along with standard data augmentations, including spatial and color transformations.

\noindent{\bfseries Setup:} We compare the synthetic image quality with DDPM \cite{ho2020denoising}, 3D-$\alpha$-WGAN \cite{kwon2019generation} and Vid2Vid \cite{wang2018video}, and utilize four segmentation models with different training strategies to demonstrate the benefit for the downstream task.

\vspace{-1em}
\subsection{Evaluate the Quality of Synthetic Image.}
\label{sec_42}

\noindent\textbf{Synthetic Dataset:} To address the limited availability of annotated 3D medical data, we used only 30 CT scans from SegTHOR (25 for training and 5 for validation) and 50 MRI scans from OASIS (40 for training and 10 for validation) to generate 110 3D thoracic CT scans and 110 3D brain MRI scans, respectively.

\vspace{-1em}
\begin{table}
\centering
\setlength{\tabcolsep}{5.2mm}
\renewcommand{\arraystretch}{0.7}{
\begin{tabular}{ccccc}
\toprule
 & \multicolumn{2}{c}{ \footnotesize{Thoracic CT} } & \multicolumn{2}{c}{ \footnotesize{Brain MRI} } \\
\cmidrule(lr){2-3} \cmidrule(lr){4-5}
 & \footnotesize{FID} $\downarrow$ & \footnotesize{LPIPS} $\uparrow$  & \footnotesize{FID} $\downarrow$ & \footnotesize{LPIPS} $\uparrow$ \\
\midrule 
\footnotesize{DDPM \cite{ho2020denoising}} & \footnotesize{\textbf{35.2}} & \footnotesize{\textbf{0.316}} & \footnotesize{\textbf{34.9}} & \footnotesize{0.298} \\
\footnotesize{3D-$\alpha$-WGAN \cite{kwon2019generation}} & \footnotesize{136.2} & \footnotesize{0.286} & \footnotesize{136.4} & \footnotesize{0.289} \\
\footnotesize{Vid2Vid \cite{wang2018video}} & \footnotesize{47.3} & \footnotesize{0.300} & \footnotesize{48.2} & \footnotesize{0.324} \\
\midrule 
\footnotesize{Ours} & \footnotesize{39.6} & \footnotesize{0.305} & \footnotesize{40.3} & \footnotesize{\textbf{0.326}} \\
\bottomrule
\end{tabular}
}
\vspace{0.5em}
\caption{Synthetic image quality comparison between baselines and ours.}
\label{t1}
\vspace{-3em}
\end{table}

We compare the fidelity and diversity of our synthetic data with DDPM \cite{ho2020denoising} (train 3 for different views), 3D-$\alpha$-WGAN \cite{kwon2019generation}, and vid2vid \cite{wang2018video} by calculating the mean Frèchet Inception Distance (FID) and Learned Perceptual Image Patch Similarity (LPIPS) from 3 different views. 


According to Table 1, our proposed method has a slightly lower FID score but a similar LPIPS score compared to DDPM. We speculate that this is because DDPM is trained on 2D images without explicit anatomical constraints and only generates 2D images. On the other hand, 3D-$\alpha$-WGAN [18], which uses much larger 3D training data (146 for thorax and 414 for brain), has significantly worse FID and LPIPS scores than our method. Moreover, our proposed method outperforms Vid2Vid, showing the effectiveness of our semantic diffusion refiner.

\vspace{-2em}

\begin{figure}
    \centering
    \includegraphics[width=\textwidth]{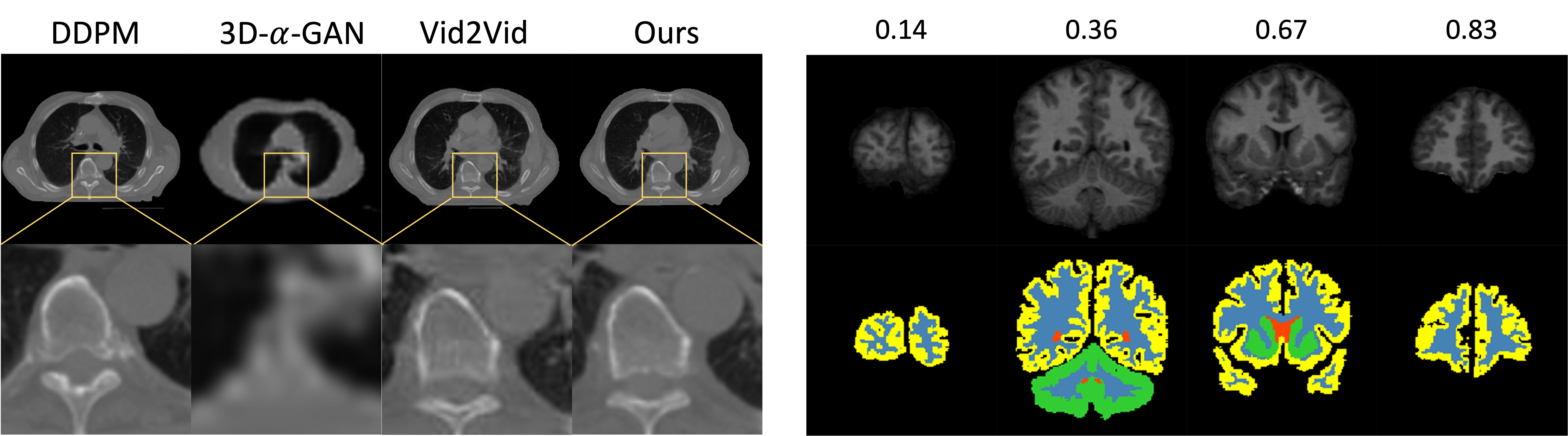}
    \caption{Our proposed method produces more anatomically accurate images compared to 3D-$\alpha$-WGAN and vid2vid, as demonstrated by the clearer organ boundaries and more realistic textures. Left: Qualitative comparison between different generative models. Right: Visualization of synthetic 3D brain MRI slices at different relative positions.}
    \label{fig:vis}
\end{figure}

\vspace{-3em}

\subsection{Evaluate the Benefits for Segmentation Task.}

We explore the benefits of synthetic data for downstream segmentation tasks by comparing Sørensen–Dice coefficient (DSC) of 4 segmentation models, including Unet2D \cite{ronneberger2015u}, UNet3D \cite{cciccek20163d}, UNETR \cite{hatamizadeh2022unetr}, and Swin-UNETR \cite{tang2022self}. In Table \ref{t2} and \ref{t3}, we utilize real training data (from SegTHOR and OASIS) and synthetic data to train the segmentation models with 5 different strategies, and test on all 3 thoracic CT datasets and 2 brain MRI datasets. In Table \ref{t4}, we aim to demonstrate whether the synthetic data can aid transfer learning with limited real finetuning data from each of the testing datasets (StructSeg-Thorax, Public-Thor and ADNI) with four training strategies. 

\vspace{-1.5em}
\begin{table}
\centering
\setlength{\tabcolsep}{0.5mm}
\renewcommand{\arraystretch}{0.65}
\renewcommand\cellset{}{
\begin{tabular}{lcccccccccccc}
\toprule
  & \multicolumn{4}{c}{ \footnotesize{SegTHOR*} } & \multicolumn{4}{c}{\footnotesize{StructSeg-Thorax} } &  \multicolumn{4}{c}{\footnotesize{Public-Thor}  } \\
\cmidrule(lr){2-5} \cmidrule(lr){6-9} \cmidrule(lr){10-13}
 & \makecell[c]{\scriptsize{ Unet }\\ \scriptsize{ 2D}} & \makecell[c]{\scriptsize{ Unet }\\ \scriptsize{ 3D}} & \scriptsize{UNETR} & \makecell[c]{\scriptsize{Swin}\\ \scriptsize{UNETR}}  
& \makecell[c]{\scriptsize{ Unet }\\ \scriptsize{ 2D}} & \makecell[c]{\scriptsize{ Unet }\\ \scriptsize{ 3D}} & \scriptsize{UNETR} & \makecell[c]{\scriptsize{Swin}\\ \scriptsize{UNETR}} 
& \makecell[c]{\scriptsize{ Unet }\\ \scriptsize{ 2D}} & \makecell[c]{\scriptsize{ Unet }\\ \scriptsize{ 3D}} & \scriptsize{UNETR} & \makecell[c]{\scriptsize{Swin}\\ \scriptsize{UNETR}}  \\
\midrule 
\scriptsize{\textbf{E2-1}} & \scriptsize{0.817} & \scriptsize{0.873} & \scriptsize{0.867} & \scriptsize{0.878} & \scriptsize{0.722} & \scriptsize{0.793} & \scriptsize{0.789} & \scriptsize{0.810} & \scriptsize{0.822} & \scriptsize{0.837} & \scriptsize{0.836} & \scriptsize{0.847}\\

\scriptsize{\textbf{E2-2}} & \scriptsize{0.815} & \scriptsize{0.846} & \scriptsize{0.845} & \scriptsize{0.854} & \scriptsize{0.736} & \scriptsize{0.788} & \scriptsize{0.788} & \scriptsize{0.803} & \scriptsize{0.786} & \scriptsize{0.838} & \scriptsize{0.814} & \scriptsize{0.842}\\

\scriptsize{\textbf{E2-3}} & \scriptsize{0.845} & \scriptsize{0.881} & \scriptsize{0.886} & \scriptsize{0.886} & \scriptsize{0.772} & \scriptsize{0.827} & \scriptsize{0.824} & \scriptsize{0.827} & \scriptsize{0.812} & \scriptsize{0.856} & \scriptsize{0.853} & \scriptsize{0.856}\\

\scriptsize{\textbf{E2-4}} & \scriptsize{\textbf{0.855}} & \scriptsize{0.887} & \scriptsize{\textbf{0.894}} & \scriptsize{\textbf{0.899}} & \scriptsize{0.775} & \scriptsize{\textbf{0.833}} & \scriptsize{\textbf{0.825}} & \scriptsize{0.833} & \scriptsize{\textbf{0.824}} & \scriptsize{0.861} & \scriptsize{0.852} & \scriptsize{\textbf{0.867}}\\

\scriptsize{\textbf{E2-5}} & \scriptsize{0.847} & \scriptsize{\textbf{0.891}} & \scriptsize{0.890} & \scriptsize{0.897} & \scriptsize{\textbf{0.783}} & \scriptsize{\textbf{0.833}} & \scriptsize{0.823} & \scriptsize{\textbf{0.835}} & \scriptsize{0.818} & \scriptsize{\textbf{0.864}} & \scriptsize{\textbf{0.858}} & \scriptsize{\textbf{0.867}}\\
\bottomrule
\end{tabular}
}
\caption{Experiment 2: DSC of different thoracic segmentation models. There are 5 training strategies, namely: \textbf{E2-1:} Training with real SegTHOR training data; \textbf{E2-2:} Training with synthetic data; \textbf{E2-3:} Training with both synthetic and real data; \textbf{E2-4:} Finetuning model from E2-2 using real training data; and \textbf{E2-5:} finetuning model from E2-3 using real training data. (* denotes the training data source.)}
\label{t2}
\end{table}
\vspace{-2em}

According to Table \ref{t2} and Table \ref{t3}, the significant DSC difference between 2D and 3D segmentation models underlines the crucial role of 3D annotated data. While purely synthetic data (\textbf{E2-2}) fails to achieve the same performance as real training data (\textbf{E2-1}), the combination of real and synthetic data (\textbf{E2-3}) improves model performance in most cases, except for Unet2D on the Public-Thor dataset. Furthermore, fine-tuning the pre-trained model with real data (\textbf{E2-4} and \textbf{E2-5}) consistently outperforms the model trained only with real data. Please refer to \textbf{Supplementary} for organ-level DSC comparisons of the Swin-UNETR model with more details.

\vspace{-1.5em}
\begin{table}
\centering
\setlength{\tabcolsep}{2.5mm}
\renewcommand{\arraystretch}{0.65}
\renewcommand\cellset{}
{
\begin{tabular}{c cccc cccc}
\toprule
  & \multicolumn{4}{c}{ \footnotesize{OASIS*} } & \multicolumn{4}{c}{\footnotesize{ADNI} } \\
\cmidrule(lr){2-5} \cmidrule(lr){6-9}
 & \makecell[c]{\scriptsize{ Unet }\\ \scriptsize{ 2D}} & \makecell[c]{\scriptsize{ Unet }\\ \scriptsize{ 3D}} & \scriptsize{UNETR} & \makecell[c]{\scriptsize{Swin}\\ \scriptsize{UNETR}}  
& \makecell[c]{\scriptsize{ Unet }\\ \scriptsize{ 2D}} & \makecell[c]{\scriptsize{ Unet }\\ \scriptsize{ 3D}} & \scriptsize{UNETR} & \makecell[c]{\scriptsize{Swin}\\ \scriptsize{UNETR}} \\

\midrule 
\scriptsize{\textbf{E2-1}} & \scriptsize{0.930}  & \scriptsize{0.951}  & \scriptsize{0.952}  & \scriptsize{0.954} & \scriptsize{0.815} & \scriptsize{0.826} & \scriptsize{0.880} & \scriptsize{0.894} \\
\scriptsize{\textbf{E2-2}} & \scriptsize{0.905}  & \scriptsize{0.936}  & \scriptsize{0.935}  & \scriptsize{0.934} & \scriptsize{0.759} & \scriptsize{0.825} & \scriptsize{0.828} & \scriptsize{0.854} \\
\scriptsize{\textbf{E2-3}} & \scriptsize{0.938} &  \scriptsize{0.953} & \scriptsize{0.953} & \scriptsize{0.955}   & \scriptsize{0.818} & \scriptsize{0.888} & \scriptsize{0.898} & \scriptsize{\textbf{0.906}} \\
\scriptsize{\textbf{E2-4}} & \scriptsize{\textbf{0.940}} &  \scriptsize{\textbf{0.955}} & \scriptsize{\textbf{0.954}} & \scriptsize{\textbf{0.956}}   & \scriptsize{\textbf{0.819}} & \scriptsize{0.891} & \scriptsize{\textbf{0.903}} & \scriptsize{0.903} \\
\scriptsize{\textbf{E2-5}} & \scriptsize{\textbf{0.940}} &  \scriptsize{0.954} & \scriptsize{\textbf{0.954}} & \scriptsize{\textbf{0.956}}   & \scriptsize{\textbf{0.819}} & \scriptsize{\textbf{0.894}} & \scriptsize{0.902} & \scriptsize{\textbf{0.906}} \\
\bottomrule
\end{tabular}
}
\vspace{0.3em}
\caption{Experiment 2: DSC of brain segmentation models. Please refer to Table \ref{t2} for detailed training strategies. (*
denotes the training data source.)
}
\label{t3}
\end{table}
\vspace{-3em}

According to Table \ref{t4}, for transfer learning, utilizing the pre-trained model (\textbf{E3-2}) leads to better performance compared to training from scratch (\textbf{E3-1}). Additionally, pretraining the model with synthetic data (\textbf{E3-3} and \textbf{E3-4}) can facilitate transfer learning to a new dataset with limited annotated data.

\vspace{-1.5em}
\begin{table}
\centering
\setlength{\tabcolsep}{6.6mm}
\renewcommand{\arraystretch}{0.65}{
\begin{tabular}{c cc c}
\toprule
 & \multicolumn{2}{c}{ \footnotesize{Thoracic CT} } & \multicolumn{1}{c}{ \footnotesize{Brain MRI} } \\
\cmidrule(lr){2-3}  \cmidrule(lr){4-4}
 & \footnotesize{StructSeg-Thorax*} & \footnotesize{Public-Thor*} & \footnotesize{ADNI*} \\
\midrule 
\scriptsize{\textbf{E3-1}} & \scriptsize{0.845} & \scriptsize{0.897} & \scriptsize{0.946} \\
\scriptsize{\textbf{E3-2}} & \scriptsize{0.865} & \scriptsize{0.901} & \scriptsize{0.948}\\
\scriptsize{\textbf{E3-3}} & \scriptsize{0.878} & \scriptsize{0.913} & \scriptsize{\textbf{0.949}} \\
\scriptsize{\textbf{E3-4}} & \scriptsize{\textbf{0.882}} & \scriptsize{\textbf{0.914}} & \scriptsize{\textbf{0.949}} \\
\bottomrule
\end{tabular}
}
\vspace{0.5em}
\caption{Experiment 3: DSC of Swin-UNETR finetuned with real dataset. There are 4 training strategies: \textbf{E3-1:} Training from scratch for each dataset using limited finetuning data; \textbf{E3-2} Finetuning the model E2-1 from experiment 2; \textbf{E3-3} Finetuning the model E2-4 from experiment 2; and \textbf{E3-4} Finetuning the model E2-5 from experiment 2. (*
denotes the finetuning data source.)
}
\label{t4}
\end{table}

\vspace{-2em}

We have included video demonstrations of the generated 3D volumetric images in the \textbf{supplementary material}, which offer a more comprehensive representation of the generated image's quality.



\vspace{-0.5em}
\section{Conclusion}
\vspace{-0.5em}

This paper introduces MedGen3D, a new framework for synthesizing 3D medical mask-image pairs. Our experiments demonstrate its potential in realistic data generation and downstream segmentation tasks with limited annotated data. Future work includes merging the image sequence generator and semantic diffusion refiner for end-to-end training and extending the framework to synthesize 3D medical images across modalities. Overall, we believe that our work opens up new possibilities for generating 3D high-quality medical images paired with masks, and look forward to future developments in this field.

\newpage
%
%
%
%

\bibliographystyle{splncs04}
\bibliography{egbib}

\begin{thebibliography}{10}
\providecommand{\url}[1]{\texttt{#1}}
\providecommand{\urlprefix}{URL }
\providecommand{\doi}[1]{https://doi.org/#1}

\bibitem{ADNI}
https://adni.loni.usc.edu/

\bibitem{structseg}
https://structseg2019.grand-challenge.org/dataset/

\bibitem{abbasi20204d}
Abbasi-Sureshjani, S., Amirrajab, S., Lorenz, C., Weese, J., Pluim, J.,
  Breeuwer, M.: 4d semantic cardiac magnetic resonance image synthesis on xcat
  anatomical model. In: Medical Imaging with Deep Learning. pp. 6--18. PMLR
  (2020)

\bibitem{abhishek2019mask2lesion}
Abhishek, K., Hamarneh, G.: Mask2lesion: Mask-constrained adversarial skin
  lesion image synthesis. In: Simulation and Synthesis in Medical Imaging: 4th
  International Workshop, SASHIMI 2019, Held in Conjunction with MICCAI 2019,
  Shenzhen, China, October 13, 2019, Proceedings. pp. 71--80. Springer (2019)

\bibitem{arjovsky2017wasserstein}
Arjovsky, M., Chintala, S., Bottou, L.: Wasserstein gan. arXiv preprint arXiv:
  Arxiv-1701.07875  (2017)

\bibitem{baur2018melanogans}
Baur, C., Albarqouni, S., Navab, N.: Melanogans: high resolution skin lesion
  synthesis with gans. arXiv preprint arXiv:1804.04338  (2018)

\bibitem{bermudez2018learning}
Bermudez, C., Plassard, A.J., Davis, L.T., Newton, A.T., Resnick, S.M.,
  Landman, B.A.: Learning implicit brain mri manifolds with deep learning. In:
  Medical Imaging 2018: Image Processing. vol. 10574, pp. 408--414. SPIE (2018)

\bibitem{cardoso2022monai}
Cardoso, M.J., Li, W., Brown, R., Ma, N., Kerfoot, E., Wang, Y., Murrey, B.,
  Myronenko, A., Zhao, C., Yang, D., et~al.: Monai: An open-source framework
  for deep learning in healthcare. arXiv preprint arXiv:2211.02701  (2022)

\bibitem{chen2021deep}
Chen, X., Sun, S., Bai, N., Han, K., Liu, Q., Yao, S., Tang, H., Zhang, C., Lu,
  Z., Huang, Q., et~al.: A deep learning-based auto-segmentation system for
  organs-at-risk on whole-body computed tomography images for radiation
  therapy. Radiotherapy and Oncology  \textbf{160},  175--184 (2021)

\bibitem{cciccek20163d}
{\c{C}}i{\c{c}}ek, {\"O}., Abdulkadir, A., Lienkamp, S.S., Brox, T.,
  Ronneberger, O.: 3d u-net: learning dense volumetric segmentation from sparse
  annotation. In: Medical Image Computing and Computer-Assisted
  Intervention--MICCAI 2016: 19th International Conference, Athens, Greece,
  October 17-21, 2016, Proceedings, Part II 19. pp. 424--432. Springer (2016)

\bibitem{dhariwal2021diffusion}
Dhariwal, P., Nichol, A.: Diffusion models beat gans on image synthesis.
  Advances in Neural Information Processing Systems  \textbf{34},  8780--8794
  (2021)

\bibitem{fernandez2022can}
Fernandez, V., Pinaya, W.H.L., Borges, P., Tudosiu, P.D., Graham, M.S.,
  Vercauteren, T., Cardoso, M.J.: Can segmentation models be trained with fully
  synthetically generated data? In: Simulation and Synthesis in Medical
  Imaging: 7th International Workshop, SASHIMI 2022, Held in Conjunction with
  MICCAI 2022, Singapore, September 18, 2022, Proceedings. pp. 79--90. Springer
  (2022)

\bibitem{fischl2012freesurfer}
Fischl, B.: Freesurfer. Neuroimage  \textbf{62}(2),  774--781 (2012)

\bibitem{goodfellow2020generative}
Goodfellow, I., Pouget-Abadie, J., Mirza, M., Xu, B., Warde-Farley, D., Ozair,
  S., Courville, A., Bengio, Y.: Generative adversarial networks.
  Communications of the ACM  \textbf{63}(11),  139--144 (2020)

\bibitem{guibas2017synthetic}
Guibas, J.T., Virdi, T.S., Li, P.S.: Synthetic medical images from dual
  generative adversarial networks. arXiv preprint arXiv:1709.01872  (2017)

\bibitem{han2018gan}
Han, C., Hayashi, H., Rundo, L., Araki, R., Shimoda, W., Muramatsu, S.,
  Furukawa, Y., Mauri, G., Nakayama, H.: Gan-based synthetic brain mr image
  generation. In: 2018 IEEE 15th international symposium on biomedical imaging
  (ISBI 2018). pp. 734--738. IEEE (2018)

\bibitem{hatamizadeh2022unetr}
Hatamizadeh, A., Tang, Y., Nath, V., Yang, D., Myronenko, A., Landman, B.,
  Roth, H.R., Xu, D.: Unetr: Transformers for 3d medical image segmentation.
  In: Proceedings of the IEEE/CVF winter conference on applications of computer
  vision. pp. 574--584 (2022)

\bibitem{ho2020denoising}
Ho, J., Jain, A., Abbeel, P.: Denoising diffusion probabilistic models.
  Advances in Neural Information Processing Systems  \textbf{33},  6840--6851
  (2020)

\bibitem{ho2022classifierfree}
Ho, J., Salimans, T.: Classifier-free diffusion guidance. arXiv preprint arXiv:
  Arxiv-2207.12598  (2022)

\bibitem{kim2021synthesis}
Kim, S., Kim, B., Park, H.: Synthesis of brain tumor multicontrast mr images
  for improved data augmentation. Medical Physics  \textbf{48}(5),  2185--2198
  (2021)

\bibitem{kwon2019generation}
Kwon, G., Han, C., Kim, D.s.: Generation of 3d brain mri using auto-encoding
  generative adversarial networks. In: Medical Image Computing and Computer
  Assisted Intervention--MICCAI 2019: 22nd International Conference, Shenzhen,
  China, October 13--17, 2019, Proceedings, Part III 22. pp. 118--126. Springer
  (2019)

\bibitem{lambert2020segthor}
Lambert, Z., Petitjean, C., Dubray, B., Kuan, S.: Segthor: Segmentation of
  thoracic organs at risk in ct images. In: 2020 Tenth International Conference
  on Image Processing Theory, Tools and Applications (IPTA). pp.~1--6. IEEE
  (2020)

\bibitem{marcus2007open}
Marcus, D.S., Wang, T.H., Parker, J., Csernansky, J.G., Morris, J.C., Buckner,
  R.L.: Open access series of imaging studies (oasis): cross-sectional mri data
  in young, middle aged, nondemented, and demented older adults. Journal of
  cognitive neuroscience  \textbf{19}(9),  1498--1507 (2007)

\bibitem{park2019semantic}
Park, T., Liu, M.Y., Wang, T.C., Zhu, J.Y.: Semantic image synthesis with
  spatially-adaptive normalization. In: Proceedings of the IEEE/CVF conference
  on computer vision and pattern recognition. pp. 2337--2346 (2019)

\bibitem{rombach2022high}
Rombach, R., Blattmann, A., Lorenz, D., Esser, P., Ommer, B.: High-resolution
  image synthesis with latent diffusion models. In: Proceedings of the IEEE/CVF
  Conference on Computer Vision and Pattern Recognition. pp. 10684--10695
  (2022)

\bibitem{ronneberger2015u}
Ronneberger, O., Fischer, P., Brox, T.: U-net: Convolutional networks for
  biomedical image segmentation. In: Medical Image Computing and
  Computer-Assisted Intervention--MICCAI 2015: 18th International Conference,
  Munich, Germany, October 5-9, 2015, Proceedings, Part III 18. pp. 234--241.
  Springer (2015)

\bibitem{subramaniam2022generating}
Subramaniam, P., Kossen, T., Ritter, K., Hennemuth, A., Hildebrand, K.,
  Hilbert, A., Sobesky, J., Livne, M., Galinovic, I., Khalil, A.A., et~al.:
  Generating 3d tof-mra volumes and segmentation labels using generative
  adversarial networks. Medical Image Analysis  \textbf{78},  102396 (2022)

\bibitem{sun2022hierarchical}
Sun, L., Chen, J., Xu, Y., Gong, M., Yu, K., Batmanghelich, K.: Hierarchical
  amortized gan for 3d high resolution medical image synthesis. IEEE journal of
  biomedical and health informatics  \textbf{26}(8),  3966--3975 (2022)

\bibitem{tang2022self}
Tang, Y., Yang, D., Li, W., Roth, H.R., Landman, B., Xu, D., Nath, V.,
  Hatamizadeh, A.: Self-supervised pre-training of swin transformers for 3d
  medical image analysis. In: Proceedings of the IEEE/CVF Conference on
  Computer Vision and Pattern Recognition. pp. 20730--20740 (2022)

\bibitem{wang2018video}
Wang, T.C., Liu, M.Y., Zhu, J.Y., Liu, G., Tao, A., Kautz, J., Catanzaro, B.:
  Video-to-video synthesis. arXiv preprint arXiv:1808.06601  (2018)

\bibitem{yan2022after}
Yan, X., Tang, H., Sun, S., Ma, H., Kong, D., Xie, X.: After-unet: Axial fusion
  transformer unet for medical image segmentation. In: Proceedings of the
  IEEE/CVF winter conference on applications of computer vision. pp. 3971--3981
  (2022)

\bibitem{you2022mine}
You, C., Dai, W., Liu, F., Su, H., Zhang, X., Staib, L., Duncan, J.S.: Mine
  your own anatomy: Revisiting medical image segmentation with extremely
  limited labels. arXiv preprint arXiv:2209.13476  (2022)

\bibitem{you2023rethinking}
You, C., Dai, W., Min, Y., Liu, F., Zhang, X., Clifton, D.A., Zhou, S.K.,
  Staib, L.H., Duncan, J.S.: Rethinking semi-supervised medical image
  segmentation: A variance-reduction perspective. arXiv preprint
  arXiv:2302.01735  (2023)

\bibitem{you2022bootstrapping}
You, C., Dai, W., Staib, L., Duncan, J.S.: Bootstrapping semi-supervised
  medical image segmentation with anatomical-aware contrastive distillation.
  arXiv preprint arXiv:2206.02307  (2022)

\bibitem{you2022class}
You, C., Zhao, R., Liu, F., Dong, S., Chinchali, S., Topcu, U., Staib, L.,
  Duncan, J.: Class-aware adversarial transformers for medical image
  segmentation. In: NeurIPS (2022)

\end{thebibliography}


\end{document}


\begin{algorithm}
\setstretch{0.7}
\caption{3D Mask Generation (Inference)}
\begin{algorithmic}[H]
\Require n: number of conditional slices, m: length of subsequence, L: length of the sequence to generate, $P_\theta$: the predicted probability distribution
\State $Z \leftarrow L - (m-1)$
\State $z \sim$ Uniform $(\{0, 1, ..., Z\})$ \Comment{Randomly pick one z as start position}
\State Initialize an empty mask sequence $\mathcal{M}$
\State $X \leftarrow \{M_z, M_{z+1}, ..., M_{z+(m-1)}\} \sim P_\theta(X_0^{P} \mid X^C = \emptyset, z)$
\State $\mathcal{M} \leftarrow \mathcal{M} \cup X$
\State //Forward Sampling
\State $z' \leftarrow z$
\While{$z' <= Z$}
    \State $z' \leftarrow z' + (m-n)$ 
    \State $X^C \leftarrow \mathcal{M}[-n:]$ \Comment{Select the last n masks as condition}
    \State $X^P \sim P_\theta(X_0^{P} \mid X^C, z')$ \Comment{Sample the following (m-n) masks}
    \State $\mathcal{M} \leftarrow \mathcal{M} \cup X^P$\Comment{Add the generated masks to the end of sequence}
    
\EndWhile
\State //Backward Sampling
\State $z' \leftarrow z$
\While{$z' >= 0$}
    \State $z' \leftarrow z' - (m-n)$
    \State $X^C \leftarrow \mathcal{M}[:n]$ \Comment{Select the first n masks as condition}
    \State $X^P \sim P_\theta(X_0^{P} \mid X^C, z')$ \Comment{Sample the previous (m-n) masks}
    \State $\mathcal{M} \leftarrow X^P \cup \mathcal{M}$ \Comment{Add the generated masks to the start of sequence}
    
\EndWhile
\State
\Return $\mathcal{M}$ \Comment{Return the generated mask sequence}
\end{algorithmic}
\end{algorithm}

\vspace{-3em}
\begin{table}[H]
\centering
 \setlength{\tabcolsep}{0.3mm}
 \renewcommand{\arraystretch}{0.75}
\renewcommand\cellset{}{
\begin{tabular}{l cccccc cccccc cccccc}
\toprule
& \multicolumn{6}{c}{SegTHOR*} & \multicolumn{6}{c}{StructSeg-Thorax} &  \multicolumn{6}{c}{Public-Thor} \\
\cmidrule(lr){2-7} \cmidrule(lr){8-13} \cmidrule(lr){14-19}
 & \scriptsize{ ll } & \scriptsize{ rl } & \scriptsize{ ht } & \scriptsize{ eso } &  \scriptsize{ tra } & \scriptsize{ spin }
 & \scriptsize{ ll } & \scriptsize{ rl } & \scriptsize{ ht } & \scriptsize{ eso } &  \scriptsize{ tra } & \scriptsize{ spin }
 & \scriptsize{ ll } & \scriptsize{ rl } & \scriptsize{ ht } & \scriptsize{ eso } &  \scriptsize{ tra } & \scriptsize{ spin }  \\
\midrule 
\scriptsize{\textbf{E2-1}} 
& \scriptsize{\textbf{0.98}} & \scriptsize{\textbf{0.99}} & \scriptsize{0.90} & \scriptsize{0.64} & \scriptsize{0.86} & \scriptsize{0.91} 
& \scriptsize{\textbf{0.95}} & \scriptsize{\textbf{0.96}} & \scriptsize{\textbf{0.91}} & \scriptsize{0.60} & \scriptsize{0.68} & \scriptsize{0.77} 
& \scriptsize{\textbf{0.97}} & \scriptsize{\textbf{0.98}} & \scriptsize{0.88} & \scriptsize{0.69} & \scriptsize{0.73} & \scriptsize{0.84}\\

\scriptsize{\textbf{E2-2}} 
& \scriptsize{\textbf{0.98}} & \scriptsize{0.98} & \scriptsize{0.91} & \scriptsize{0.55} & \scriptsize{0.83} & \scriptsize{0.89} 
& \scriptsize{0.94} & \scriptsize{0.95} & \scriptsize{0.89} & \scriptsize{0.54} & \scriptsize{0.63} & \scriptsize{0.86} 
& \scriptsize{\textbf{0.97}} & \scriptsize{\textbf{0.98}} & \scriptsize{0.88} & \scriptsize{0.68} & \scriptsize{0.69} & \scriptsize{0.85}\\

\scriptsize{\textbf{E2-3}} 
& \scriptsize{\textbf{0.98}} & \scriptsize{0.98} & \scriptsize{0.92} & \scriptsize{0.64} & \scriptsize{0.87} & \scriptsize{0.93} 
& \scriptsize{\textbf{0.95}} & \scriptsize{0.95} & \scriptsize{0.90} & \scriptsize{0.64} & \scriptsize{0.65} & \scriptsize{\textbf{0.87}} 
& \scriptsize{\textbf{0.97}} & \scriptsize{\textbf{0.98}} & \scriptsize{0.88} & \scriptsize{0.73} & \scriptsize{0.72} & \scriptsize{\textbf{0.86}}\\

\scriptsize{\textbf{E2-4}} 
& \scriptsize{\textbf{0.98}} & \scriptsize{\textbf{0.99}} & \scriptsize{\textbf{0.93}} & \scriptsize{\textbf{0.67}} & \scriptsize{\textbf{0.90}} & \scriptsize{\textbf{0.94}} 
& \scriptsize{\textbf{0.95}} & \scriptsize{\textbf{0.96}} & \scriptsize{0.90} & \scriptsize{0.64} & \scriptsize{\textbf{0.69}} & \scriptsize{\textbf{0.87}} 
& \scriptsize{\textbf{0.97}} & \scriptsize{\textbf{0.98}} & \scriptsize{\textbf{0.90}} & \scriptsize{0.74} & \scriptsize{0.75} & \scriptsize{\textbf{0.86}}\\

\scriptsize{\textbf{E2-5}} 
& \scriptsize{\textbf{0.98}} & \scriptsize{\textbf{0.99}} & \scriptsize{0.92} & \scriptsize{\textbf{0.67}} & \scriptsize{0.88} & \scriptsize{\textbf{0.94}}
& \scriptsize{\textbf{0.95}} & \scriptsize{\textbf{0.96}} & \scriptsize{0.90} & \scriptsize{\textbf{0.67}} & \scriptsize{\textbf{0.69}} & \scriptsize{0.86} 
& \scriptsize{\textbf{0.97}} & \scriptsize{\textbf{0.98}} & \scriptsize{0.89} & \scriptsize{\textbf{0.75}} & \scriptsize{\textbf{0.77}} & \scriptsize{0.85}\\
\midrule
& \multicolumn{3}{c}{\scriptsize{ll: left lung}} &  \multicolumn{3}{c}{\scriptsize{rl: right lung}} &  \multicolumn{3}{c}{\scriptsize{ht: heart}} &  \multicolumn{3}{c}{\scriptsize{eso: esophagus}} & \multicolumn{3}{c}{\scriptsize{tra: trachea}} &  \multicolumn{3}{c}{\scriptsize{spin: spinal cord}} \\
\bottomrule
\end{tabular}
 }
\caption{Experiment 2: Organ-level DSC Comparison of Swin-UNETR for thoracic site. Please refer to \textbf{Table 2} in \textbf{main} submission for detailed strategies.}

\end{table}
\vspace{-3em}
\begin{table}[H]
\centering
\setlength{\tabcolsep}{1.5mm}
\renewcommand{\arraystretch}{0.65}
\renewcommand\cellset{}{
\begin{tabular}{l cccc cccc}
\toprule
  & \multicolumn{4}{c}{ OASIS* } & \multicolumn{4}{c}{ADNI } \\
\cmidrule(lr){2-5} \cmidrule(lr){6-9} 
 & \scriptsize{ Cortex } & \makecell[c]{\scriptsize{ Subcortical }\\ \scriptsize{ Gray}} & \makecell[c]{\scriptsize{ White }\\ \scriptsize{ Matter}} & \scriptsize{ CSF }
 & \scriptsize{ Cortex } & \makecell[c]{\scriptsize{ Subcortical }\\ \scriptsize{ Gray}} & \makecell[c]{\scriptsize{ White }\\ \scriptsize{ Matter}} & \scriptsize{ CSF } \\

\midrule 
\scriptsize{\textbf{E2-1}} & \scriptsize{0.941}  & \scriptsize{0.965}  & \scriptsize{0.970}  & \scriptsize{0.941} & \scriptsize{0.884} & \scriptsize{0.856} & \scriptsize{0.929} & \scriptsize{0.908} \\
\scriptsize{\textbf{E2-2}} & \scriptsize{0.922}  & \scriptsize{0.951}  & \scriptsize{0.958}  & \scriptsize{0.909} & \scriptsize{0.847} & \scriptsize{0.841} & \scriptsize{0.910} & \scriptsize{0.818} \\
\scriptsize{\textbf{E2-3}} & \scriptsize{0.942} &  \scriptsize{0.964} & \scriptsize{0.970} & \scriptsize{0.944}   & \scriptsize{\textbf{0.895}} & \scriptsize{\textbf{0.876}} & \scriptsize{\textbf{0.932}} & \scriptsize{\textbf{0.925}} \\
\scriptsize{\textbf{E2-4}} & \scriptsize{\textbf{0.943}} &  \scriptsize{0.965} & \scriptsize{\textbf{0.971}} & \scriptsize{0.946} & \scriptsize{0.888} & \scriptsize{0.870} & \scriptsize{0.931} & \scriptsize{0.924} \\
\scriptsize{\textbf{E2-5}} & \scriptsize{\textbf{0.943}} &  \scriptsize{\textbf{0.966}} & \scriptsize{\textbf{0.971}} & \scriptsize{\textbf{0.947}}  & \scriptsize{\textbf{0.895}} & \scriptsize{0.875} & \scriptsize{0.931} & \scriptsize{0.919} \\
\bottomrule
\end{tabular}
}
\vspace{0.5em}
\caption{Experiment 2: Organ-level DSC Comparison of Swin-UNETR for brain site.}

\end{table}

\vspace{-3em}

\begin{figure}
\centering
\includegraphics[width=\textwidth]{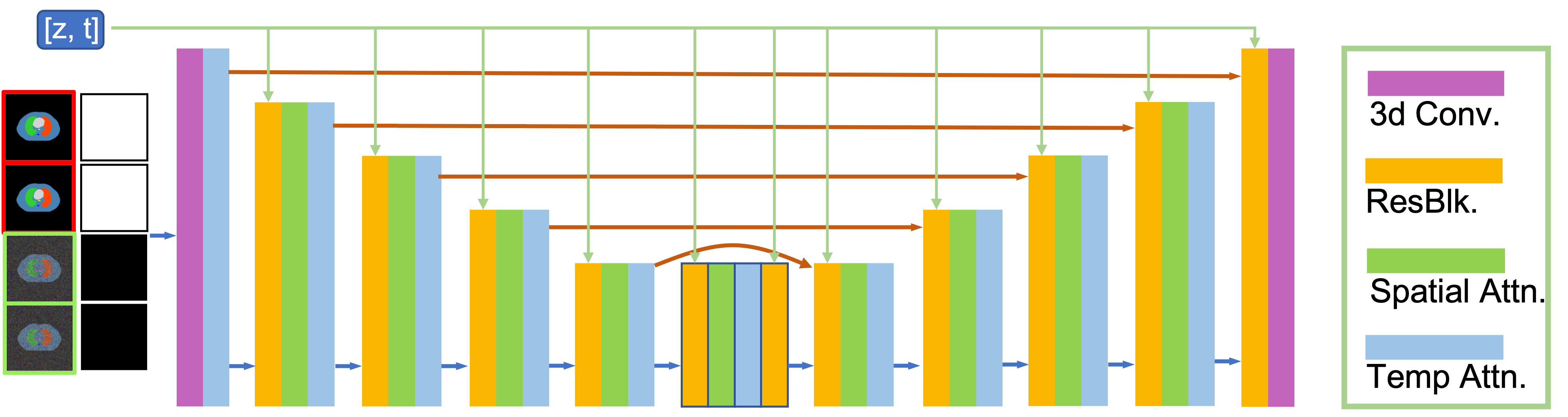}
\vspace{-1em}
\caption{Network Structure of MC-DPM.} \label{supp_p3}
\end{figure}
\vspace{-5em}
\begin{figure}
\centering
\includegraphics[width=\textwidth]{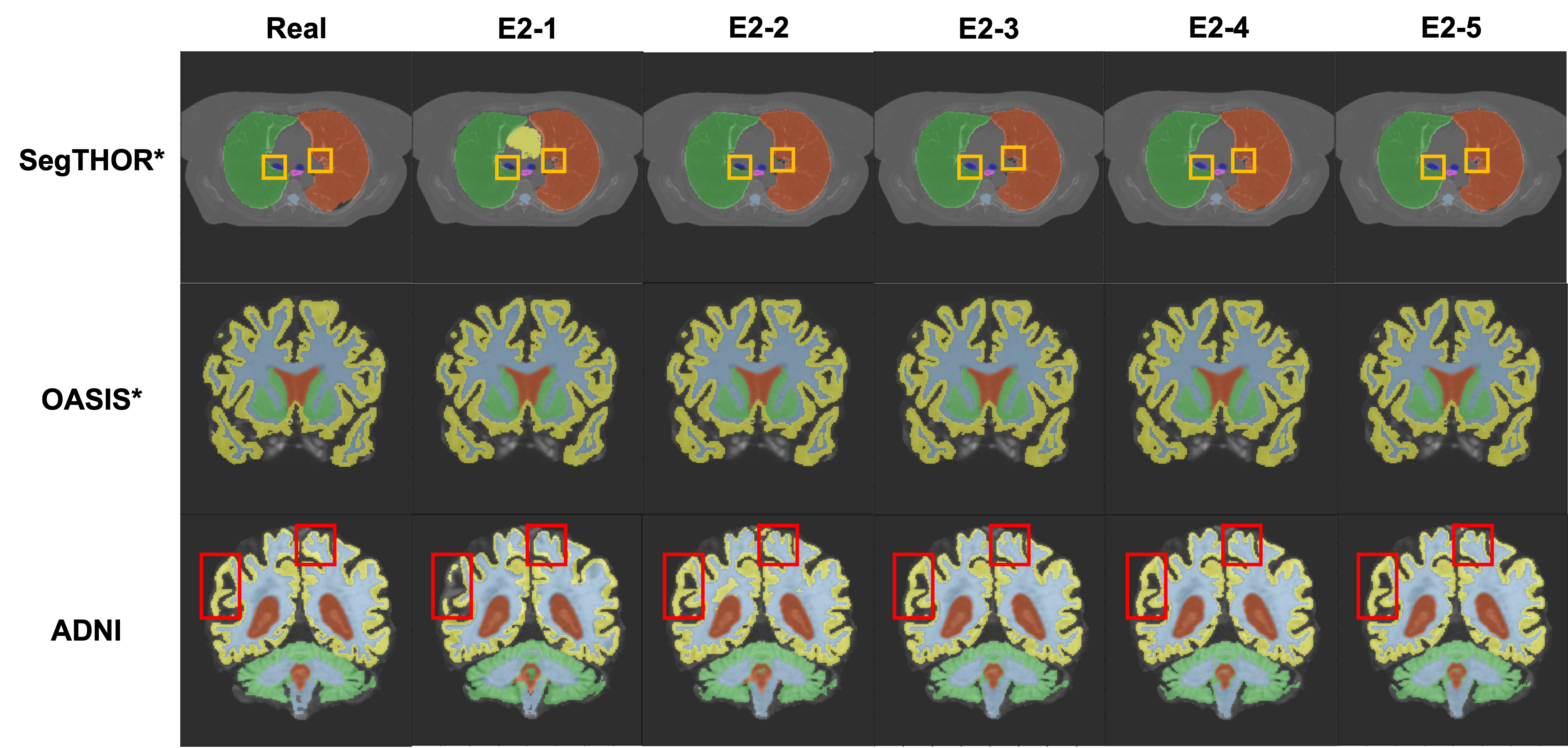}
\vspace{-2em}
\caption{Experiment 2: Qualitative comparison of different training strategies using Swin-UNETR. For brain segmentation, improvements brought by synthetic data are more evident when training and testing data come from different datasets (ADNI: train on OASIS, test on ADNI) rather than same dataset (OASIS*).} 
\label{supp_p1}
\end{figure}
\vspace{-4em}
\begin{figure}
\centering
\includegraphics[width=0.8\textwidth]{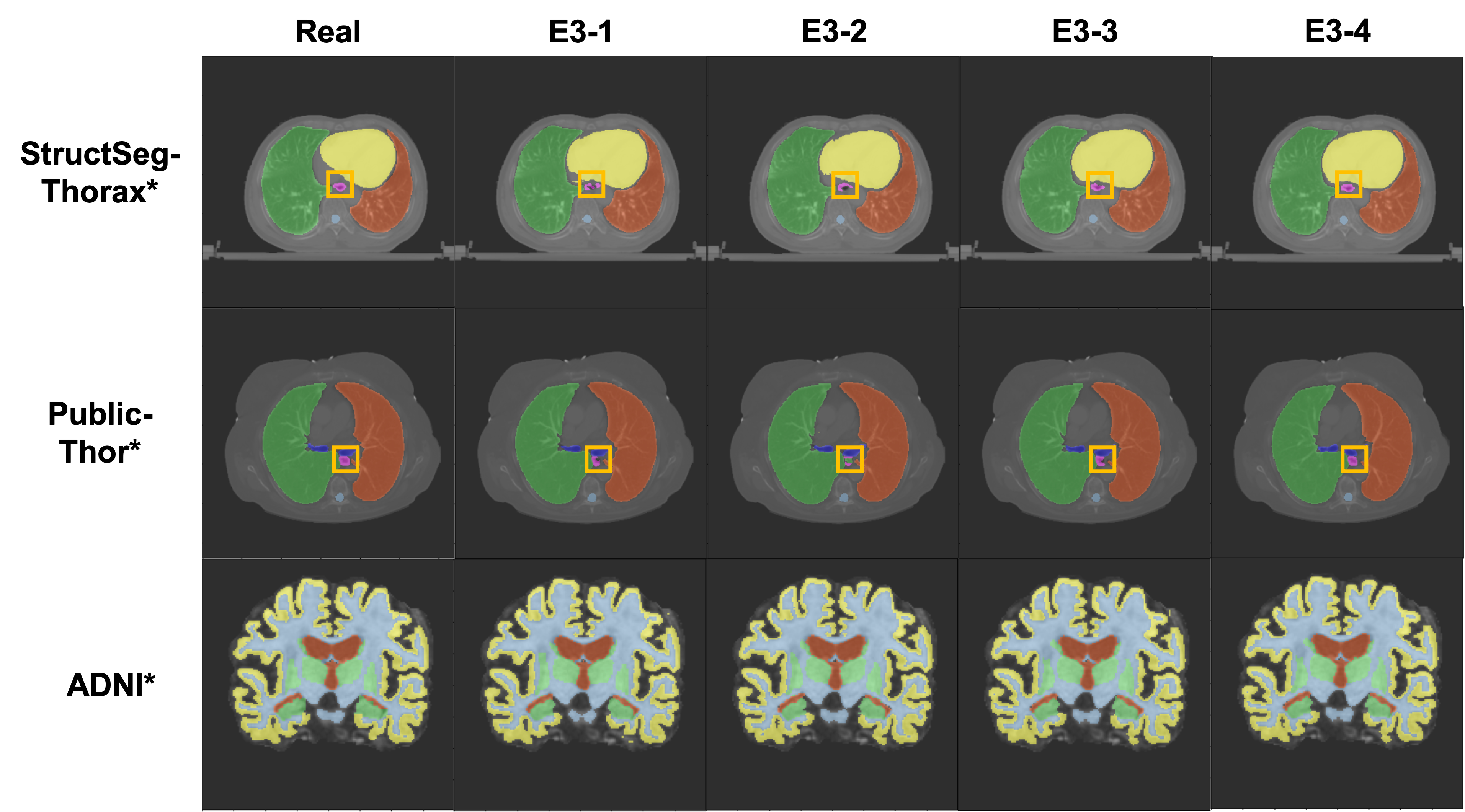}
\vspace{-1em}
\caption{Experiment 3: Qualitative comparison of different finetuning strategies using Swin-UNETR. Please refer to \textbf{Table 4} in \textbf{main} submission for detailed strategies.} \label{supp_p2}
\end{figure}